\documentclass[12pt,epsf]{article}
\usepackage{amsmath,amssymb,graphicx,epsfig,latexsym,cancel}
\numberwithin{equation}{section}
\newcommand{\be}{\begin{equation}}
\newcommand{\ee}{\end{equation}}

\newcommand{\beqa}{\begin{eqnarray}}
\newcommand{\eeqa}{\end{eqnarray}}
\newcommand{\nn}{\nonumber}

\usepackage{amsmath,amssymb,graphicx,epsfig,latexsym}

\def\boxit#1{\vbox{\hrule\hbox{\vrule\kern8pt
\vbox{\hbox{\kern8pt}\hbox{\vbox{#1}}\hbox{\kern8pt}}
\kern8pt\vrule}\hrule}}
\def\mathboxit#1{\vbox{\hrule\hbox{\vrule\kern8pt\vbox{\kern8pt
\hbox{$\displaystyle #1$}\kern8pt}\kern8pt\vrule}\hrule}}

\def\IB{\relax\hbox{$\inbar\kern-.3em{\rm B}$}}
\def\IC{\relax\hbox{$\inbar\kern-.3em{\rm C}$}}
\def\ID{\relax\hbox{$\inbar\kern-.3em{\rm D}$}}
\def\IE{\relax\hbox{$\inbar\kern-.3em{\rm E}$}}
\def\IF{\relax\hbox{$\inbar\kern-.3em{\rm F}$}}
\def\IG{\relax\hbox{$\inbar\kern-.3em{\rm G}$}}
\def\IGa{\relax\hbox{${\rm I}\kern-.18em\Gamma$}}
\def\IH{\relax{\rm I\kern-.18em H}}
\def\IK{\relax{\rm I\kern-.18em K}}
\def\IL{\relax{\rm I\kern-.18em L}}
\def\IP{\relax{\rm I\kern-.18em P}}
\def\IR{\relax{\rm I\kern-.18em R}}
\def\IZ{\relax\ifmmode\mathchoice
{\hbox{\cmss Z\kern-.4em Z}}{\hbox{\cmss Z\kern-.4em Z}}
{\lower.9pt\hbox{\cmsss Z\kern-.4em Z}} {\lower1.2pt\hbox{\cmsss
Z\kern-.4em Z}}\else{\cmss Z\kern-.4em Z}\fi}

\def\II{\relax{\rm I\kern-.18em I}}

\def\CD {{\cal D}}

\def\CP {{\cal P}}

\begin{document}

\setlength{\baselineskip}{7mm}
\begin{titlepage}

\begin{flushright}

{\tt NRCPS-HE-2-2014} \\

\end{flushright}

\vspace{1cm}
\begin{center}

{\Large \it Extension of Chern-Simons Forms

\vspace{0,3cm}

} 

\vspace{1cm}
{\sl Spyros Konitopoulos
 and  George Savvidy

\bigskip
\centerline{${}^+$ \sl Demokritos National Research Center, Ag. Paraskevi,  Athens, Greece}
\bigskip
}
\end{center}
\vspace{25pt}

\centerline{{\bf Abstract}}

\vspace{10pt}

\noindent

We investigate  metric independent, gauge invariant and closed forms in the generalized YM theory.
These forms are polynomial on the corresponding fields strength tensors - curvature forms
and are analogous to the Pontryagin-Chern densities in the YM gauge theory.
The corresponding secondary characteristic classes have been expressed in integral form
in analogy with the Chern-Simons form. Because they are not unique,
the secondary forms can be dramatically simplified by the addition of
properly chosen differentials of one-step-lower-order forms.
Their gauge variation can also be found yielding the potential anomalies in the gauge field theory.

\begin{center}
\end{center}

\vspace{150 pt}


\end{titlepage}

\newpage


\section{{\it Introduction}}

The chiral anomalies, Abelian and non-Abelian \cite{Adler:1969gk,Bell:1969ts,Bardeen:1969md,Wess:1971yu,
Frampton:1983nr,Zumino:1983ew,Stora:1983ct,Faddeev:1984jp,Faddeev:1985iz,LBL-16443,
Manes:1985df,Treiman:1986ep,Faddeev:1987hg,AlvarezGaume:1985ex}, can be derived
by a differential geometric method without having to evaluate
Feynman diagrams. Indeed, the non-Abelian anomaly  in $(2n-2)$-dimensional space-time may be obtained  from
the Abelian anomaly in $2n$ dimensions by a series of reduction (transgression) steps
\cite{Zumino:1983ew,Stora:1983ct,Faddeev:1984jp,Faddeev:1985iz,LBL-16443,Manes:1985df,Treiman:1986ep,Faddeev:1987hg,AlvarezGaume:1985ex}.
The $U_A(1)$  gauge anomaly is given by the  Pontryagin-Chern-Simons
 $2n$-form
\cite{Zumino:1983ew,Stora:1983ct,Faddeev:1984jp,Faddeev:1985iz,LBL-16443,Manes:1985df,Treiman:1986ep,Faddeev:1987hg,AlvarezGaume:1985ex}:
\be\label{abeliananom}
d*J^A ~\propto ~\CP_{2n}=Tr(G^n)= d ~\omega_{2n-1},
\ee
where $\omega_{2n-1}$ is  the  Chern-Simons form in $2n-1$ dimensions
\cite{Zumino:1983ew,Stora:1983ct,Treiman:1986ep}:
\be\label{integralformforAbelian}
\omega_{2n-1}(A)= n \int^1_0 d t ~ Str(A,G^{n-1}_t),
\ee
$G=dA +A^2$ is the 2-form  Yang-Mills (YM) field-strength tensor
of the 1-form vector field\footnote{$L^a$ are the generators of the Lie algebra.}
$A = -ig A^{a}_{\mu} L^a dx^{\mu}$ and
$G_{t}= t G +(t^2-t)A^{2}$. The non-Abelian anomaly
\cite{Adler:1969gk,Bell:1969ts,Bardeen:1969md,Wess:1971yu,
Frampton:1983nr} can be obtained by the
gauge variation of $\omega_{2n-1}$
\cite{Zumino:1983ew,Stora:1983ct,Faddeev:1984jp,Faddeev:1985iz,LBL-16443,
Treiman:1986ep,Faddeev:1987hg,AlvarezGaume:1983cs,AlvarezGaume:1985yb,AlvarezGaume:1985ex}:
\be\label{definition}
 \delta    \omega_{2n-1} = d \omega^1_{2n-2}~,
\ee
where the $(2n-2)$-form has the following integral representation \cite{Zumino:1983ew,Stora:1983ct}:
\beqa\label{celebratedanomaly}
\omega^{1}_{2n-2}(\xi, A)=n(n-1) \int^{1}_{0} d t (1-t)~
Str \left(  d\xi ,   A , ~G^{n-2}_{t}  \right)\, .
\eeqa
Here $\xi = \xi^a L_a$ is a scalar gauge parameter and $Str$ denotes a symmetrized trace
\footnote {In this article we
shall use the symmetrized trace
\beqa\label{symmetric}
Str(A_{1}, A_{2}, ..., A_{n})\equiv {1\over n!}\sum_{(i_{1}, ..., i_{n})} (A_{i_{1}}
A_{i_{2}}...A_{i_{n}}),
\eeqa
where the sum
is over all permutations. Its properties are described in the Appendix B of the article \cite{LBL-16443}.
}.
The covariant divergence of the non-Abelian
left and right handed currents is given by this $(2n-2)$-form.

In recent articles \cite{arXiv:1001.2808,Antoniadis:2012ep,Antoniadis:2013jja} the authors
found closed invariant forms similar to the Pontryagin-Chern-Simons forms in
non-Abelian tensor gauge field theory \cite{Savvidy:2005fi,Savvidy:2005zm,Savvidy:2005ki}.
The first series of closed invariant forms are defined in  $\CD=2n+4 $ dimensions
and are given by the expression
\be\label{generalized}
\Phi_{2n+4} =tr (G_4 G^n)= Str(G_4, G^{n}) = d \psi_{2n+3}~,
\ee
where the corresponding secondary $(2n+3)$-form $\psi_{2n+3}$ is  in $\CD=2n+3 $ dimensions
\be\label{generalized1}
\psi_{2n+3} = Str(A_{3}, G^{n})
\ee
and $G_4 =d A_3 + \{ A, A_3\}$\footnote{In the Appendix one can find the definition of tensor
gauge fields and the corresponding
curvature forms.}. It turns out that the introduction of Str in the above equations
leads to very crucial simplifications in all our subsequent derivations.
For compact notation, when some of the entries
of Str are the same, we write them in power form.
The second series of forms is defined in $\CD=2n+6$ dimensions \cite{Antoniadis:2013jja}:
\be
\Xi_{2n+6}=Str (G_6, G^{n} ) +n Str (G^2_{4},  G^{n-1}  )  =d \phi_{2n+5}.
\ee
The general expression for the secondary $(2n+5)$-form $\phi_{2n+5}$ will be
constructed in this article. The third series of invariant closed forms
found in this article $\Upsilon_{2n+8}$ in
$\CD=2n+8$ dimensions is
\beqa\label{thethird}
\Upsilon_{2n+8}&=&Str (G_{8}, G^{n} )+3n Str (G_{4}, G_{6},
G^{n-1} )+n(n-1) Str (G^3_{4},
G^{n-2} )= d\rho_{2n+7}.
\eeqa
Its secondary form $\rho_{2n+7}$ will be presented in the next sections.

All forms $\Phi_{2n+4} $, $\Xi_{2n+6}$ and
$\Upsilon_{2n+8}$ are analogous to the   Pontryagin-Chern-Simons densities
$\CP_{2n}$  in the YM gauge theory (\ref{abeliananom}) in the sense that they
are {\it gauge invariant, closed and metric independent}.
Our aim is to investigate this rich class of topological invariants of extended gauge theory
as well as to find out potential gauge anomalies performing transgressions
analogous to (\ref{abeliananom}) and (\ref{definition}):
\be
\CP_{2n}~ \Rightarrow ~\omega_{2n-1} ~\Rightarrow ~\omega^{1}_{2n-2}.
\ee
Therefore we shall perform the following transgressions:
\beqa\label{cohomology}
&\Phi_{2n+4} ~ \Rightarrow ~\psi_{2n+3} ~\Rightarrow ~\psi^{1}_{2n+2} ,~~~\nn\\
&\Xi_{2n+6}          ~ \Rightarrow ~ \phi_{2n+5}   ~\Rightarrow ~\phi^{1}_{2n+4} ,~~~\\
&\Upsilon_{2n+8}  ~ \Rightarrow ~\rho_{2n+7} ~\Rightarrow ~\rho^{1}_{2n+6}.\nn
\eeqa
We shall find explicit expressions for these primary invariants in terms of
higher order polynomials of the curvature forms on a vector bundle.
The most difficult challenge will be the evaluation and differentiation of the
very complicated noncommutative polynomial expressions as well as the search of
the most simple expressions for the secondary forms.  The secondary forms are not uniquely
defined. Indeed, the secondary form $\psi_{2n+3}$ is defined modulo the exterior derivative
of an arbitrary $(2n +2)$-form  $\psi_{2n+3} \backsim \psi_{2n+3} + d \alpha_{2n+2}$,
the form $\phi_{2n+5}$ modulo the exterior derivative
of a $(2n+4)$-form $\phi_{2n+5} \backsim \phi_{2n+5} + d \beta_{2n+4}$
and the form $\rho_{2n+7}$ modulo the exterior derivative
of a $(2n+6)$-form $\rho_{2n+7} \backsim \rho_{2n+7} + d \gamma_{2n+6} $.
When the difference of two closed forms is an exact form, they are said
to be cohomologous to each other.
Therefore the problem is
to find out the most simple representatives in the set of equivalence classes.
Conveniently chosen exact forms will dramatically simplify the expressions.
These problems will be solved by using properties of symmetrized
traces (\ref{symmetric}) defined in \cite{LBL-16443}.

In Section~2 we shall present a general construction and analysis of the primary forms
$\Phi_{2n+4} $ and $\Xi_{2n+6}$, their secondary forms $\psi_{2n+3}$ and $\phi_{2n+5}$
and the corresponding anomalies represented by $\psi^{1}_{2n+2}$ and $\phi^{1}_{2n+4}$.
The material of this section is not completely new, but the alternative derivation
in terms of symmetrized traces will allow to extend the results to the higher-dimensional
forms $\Upsilon_{2n+8}$. In Section~3 we shall derive the
explicit expressions for the primary form $\Upsilon_{2n+8}$,
written in terms of the symmetrized traces (\ref{symmetric}), starting from the
low-dimensional forms listed in \cite{Antoniadis:2013jja}.  Next, we shall  find
the secondary forms $\rho_{2n+7}$ and the corresponding gauge anomalies
$\rho^{1}_{2n+6}$ associated with each of the independent gauge transformations $\delta_{\xi},
\delta_{\zeta_{2}},\delta_{\zeta_{4}}, \delta_{\zeta_{6}}$.
In the conclusion we summarize the primary and secondary invariant forms constructed
in the article.    In the Appendix we present useful formulas
for the gauge transformations of the fields, the corresponding Bianchi identities
and a one-parameter deformation of fields generalizing deformation of
\cite{Zumino:1983ew,Stora:1983ct}.

\section{{\it Gauge and Metric Independent Forms}}

We shall start by deriving the already known results for the form $\Phi_{2n+4}$ using the
properties of the symmetrized traces. This approach will allow to extend the
derivation to  more complicated cases. Indeed, the form can be represented in terms of
a symmetrized trace as \cite{Antoniadis:2012ep,Antoniadis:2013jja}
\beqa\label{tensoranomaly}
\Phi_{2n+4}&=&Str\Big(G_{4}, G^{n}\Big)~
\eeqa
so that its gauge invariance with respect to the standard-scalar gauge transformations $\delta_{\xi}$
and the tensor gauge transformations $\delta_{\zeta_{2}}$
can be easily checked:
\beqa
\delta_{\xi}\Phi_{2n+4}&=&Str\Big(\delta_{\xi}G_{4}, G^{n}\Big)
+nStr\Big(G_{4}, \delta_{\xi}G, G^{n-1}\Big)=\nn \\
&=&Str\Big([G_{4}, \xi], G^{n}\Big)+nStr\Big(G_{4}, [G, \xi],
G^{n-1}\Big)=0,\nn\\
\delta_{\zeta_{2}}\Phi_{2n+4}&=&Str\Big(\delta_{\zeta_{2}}G_{4}, G^{n}\Big)
=Str\Big([G, \zeta_{2}], G^{n}\Big)=\nn \\
&=&{1\over n+1}\bigg[
Str\Big([G, \zeta_{2}], G^{n}\Big)+ ... +Str\Big(G^{n},
[G, \zeta_{2}]\Big)
\bigg]=0.
\eeqa
On the last steps we used the identity (B.10) of the Appendix
B of the article \cite{LBL-16443}.
Our next step is to check that $\Phi_{2n+4}$ is a closed form. Indeed,
\beqa
d\Phi_{2n+4}(A,A_3)=Str\Big(DG_{4}, G^{n}\Big)+
n Str\Big(G_{4}, DG, G^{n-1}\Big) =Str\Big([G, A_{3}],
G^{n}\Big)=\nn \\
= {1\over n+1}\bigg[
Str\Big([G, A_{3}], G^{n}\Big)+Str\Big(G, [G, A_{3}], G^{n-1}\Big)+
...+Str\Big(G^{n}, [G, A_{3}]\Big)\bigg]=0, \nn
\eeqa
where we used the Bianchi identity $DG=0$.
To find out the secondary form we shall use a one-parameter deformation
of the gauge potentials: \cite{Zumino:1983ew,Antoniadis:2012ep,Antoniadis:2013jja}
$$
A_{t}= t A,~~~A_{3t}= t A_3,~~~A_{5t}= t A_5,~~~A_{7t}= t A_7,~~~A_{9t}= t A_9,....,
$$
defined in the Appendix (\ref{Zumino pontentials - connections}),
(\ref{derivatives of zumino tensors}), take the derivative and
employ  (B.13) of  \cite{LBL-16443}:
\beqa
{d\over dt}\Phi_{2n+4}(A_{t}, A_{3t})&=&
Str\Big({dG_{4t}\over dt}, G_{t}^{n}\Big)+n Str\Big(G_{4t}, {dG_{t}\over dt},
G_{t}^{n-1}\Big)=\nn \\
&=& Str\Big(D_{t}A_{3}+t\{A, A_{3}\},  G_{t}^{n}\Big)+
n Str\Big(G_{4t}, D_{t}A, G_{t}^{n-1}\Big)=\nn \\
&=&dStr\Big(A_{3},  G_{t}^{n}\Big)+
ndStr\Big(G_{4t}, A,  G_{t}^{n-1}\Big)+\nn \\
&&+Str\Big(\{A, A_{3t}\},  G_{t}^{n}\Big)-
nStr\Big([G_{t}, A_{3t}], A,  G_{t}^{n-1}\Big)=\nn \\
&=&d \bigg\{
Str\Big(A_{3},  G_{t}^{n}\Big)+
nStr\Big(G_{4t}, A,  G_{t}^{n-1}\Big)
\bigg\}.
\eeqa
Integrating the above equation over the parameter t in the interval $[0,1]$ we get
the integral representation of the secondary form
\beqa\label{integral}
\psi_{2n+3}=\int_{0}^{1}dt\; \bigg[Str\Big(A_{3},  G_{t}^{n}\Big)+
nStr\Big(G_{4t},  A, G_{t}^{n-1}\Big)\bigg].
\eeqa
This secondary form is not unique. It can be modified by the addition of the differential
of a one-step-lower-order form $d \alpha_{2n+2}$. Two closed forms which differ by an exact form are said to be cohomologous
to each other. The integral on the right hand side of the equation looks complicated, but
if we add a properly chosen exact form $d \alpha_{2n+2}$, then it
will be dramatically simplified. Let us take it in the following form:
$$
\alpha_{2n+2}= - n\int_{0}^{1}dt\; Str \Big(A_{3t}, A,  G_{t}^{n-1}
\Big).
$$
Then we get:
\beqa
\psi_{2n+3} &\backsim& \psi_{2n+3}+ d \alpha_{2n+2} =\psi_{2n+3}-
n\int_{0}^{1}dt\; Str\Big(D_{t}A_{3t}, A,  G_{t}^{n-1}\Big)+\nn\\
&+&
n\int_{0}^{1}dt\; Str\Big(A_{3t}, D_{t}A,  G_{t}^{n-1}\Big)-
 n(n-1)\int_{0}^{1}dt\; Str\Big(A_{3t}, A, D_{t}G_{t},
 G_{t}^{n-2}\Big), \nn
\eeqa
where $D_{t}G_{t}=0$
\footnote {The symbol "$\backsim$" denotes the cohomology relation between the two forms.}
. Using
the relations  (\ref{field-strength tensors}), (\ref{derivatives of zumino tensors})
and the integral representation
(\ref{integral}), we have:
\beqa
&&\psi_{2n+3} \backsim \psi_{2n+3}-
n\int_{0}^{1}dt\; Str\Big(G_{4t}, A,  G_{t}^{n-1}\Big)+
n\int_{0}^{1}dt\; Str\Big(A_{3t}, {\partial G_{t}\over \partial t},  G_{t}^{n-1}\Big)=\nn \\
&&= \int_{0}^{1}dt\; \bigg\{
Str \Big(A_{3},  G_{t}^{n}\Big)+
nStr\Big(A_{3t}, {\partial G_{t}\over \partial t},  G_{t}^{n-1}\Big)\bigg\}=
\int_{0}^{1}dt\; {\partial \over \partial t}Str\Big(A_{3t},  G_{t}^{n}\Big)=
Str\Big(A_{3},  G^{n}\Big).\nn \\
\eeqa
Thus, the secondary form gets the following compact form
\beqa\label{ready}
\psi_{2n+3}=Str \Big(A_{3},  G^{n}\Big).
\eeqa
The secondary forms (\ref{integral}) and (\ref{ready}) are representatives of the same cohomology
class, because their difference is an exact form $d \alpha_{2n+2}$, but, as one can see (\ref{ready}),
has a much more simple expression. Using (\ref{field-strength tensors}), (\ref{Bianchi Identities})
we can verify that $d\psi_{2n+3}=\Phi_{2n+4}$:
\beqa
dStr \bigg(A_{3}, G^{n} \bigg)&=&
Str \bigg(DA_{3},  G^{n}\bigg)-
n Str \bigg(A_{3}, DG,  G^{n-1} \bigg) = Str\bigg(G_{4},  G^{n}\bigg).\nn
\eeqa
The form (\ref{ready}) allows to find the potential  anomalies of the theory, by the
following transgression steps. The gauge invariance of the primary form $\Phi_{2n+4}$ means that
$\delta \Phi_{2n+4}=d(\delta \psi_{2n+3})=0$. By employing the Poincare's lemma it follows that
$\delta \psi_{2n+3}=d\psi^{1}_{2n+2}$, where $\psi^{1}_{2n+2}$ is the potential anomaly.
Thus, in order to proceed we have to calculate the gauge variation of the secondary form with
respect to the gauge transformations $\delta_{\xi}$ and $\delta_{\zeta_{2}}$.
We have\footnote{The identity (B.10) of the Appendix
B   \cite{LBL-16443} should be used.}
\beqa
\delta_{\xi}\psi_{2n+3}&=&Str\Big(\delta_{\xi}A_{3},  G^{n}\Big)+
nStr\Big(A_{3}, \delta_{\xi}G,  G^{n-1}\Big)=\nn \\
&=&Str\Big([A_{3}, \xi],  G^{n}\Big)+
nStr\Big(A_{3}, [G, \xi],  G^{n-1}\Big)=0,
\eeqa
that is, the secondary form is gauge invariant with respect to standard-scalar gauge transformations
$\delta_{\xi}$ and therefore there are no gauge anomalies associated with the scalar  gauge
transformations. But with respect to the tensor gauge transformations $\delta_{\zeta_{2}}$  there are anomalies
\beqa
\delta_{\zeta_{2}}\psi_{2n+3}&=&Str\Big(\delta_{\zeta_{2}}A_{3},  G^{n-1}\Big)
=Str\Big(D\zeta_{2},  G^{n}\Big)=dStr\Big(\zeta_{2},  G^{n}\Big).
\eeqa
Therefore the anomaly is
\beqa\label{amomaly0}
\psi_{2n+2}^{(1)}(\zeta_{2}, A)&=&Str\Big(\zeta_{2},  G^{n}\Big).
\eeqa
{\it In summary, we have the expressions (\ref{tensoranomaly}) for primary form $\Phi_{2n+4}$,
the expression (\ref{ready}) for the secondary form $\psi_{2n+3}$ and (\ref{amomaly0}) for the anomaly.}
\\

We shall now move to the next primary form $\Xi_{2n+6}$, which can be written in terms of
symmetrized traces as \cite{Antoniadis:2013jja}:
\be\label{tensoranomaly2}
\Xi_{2n+6}=Str\bigg(G_6, G^{n}\bigg)+nStr\bigg( G^2_{4},  G^{n-1} \bigg).
\ee
Each term of this expression is independently gauge invariant. Indeed,
\beqa
\delta_{\xi}Str\Big(G_{6},  G^{n}\Big)&=&
Str\Big(\delta_{\xi}G_{6},  G^{n}\Big)+nStr\Big( G_{6}, \delta_{\xi} G,  G^{n-1}\Big)=\nn \\
&=&Str\Big([G_{6}, \xi],   G^{n}\Big)+nStr\Big( G_{6}, [G, \xi],  G^{n-1}\Big)=0,
\eeqa
and for the second term we shall get
\beqa
\delta_{\xi}Str\Big( G^2_{4},  G^{n-1}\Big)&=&
2Str\Big(\delta_{\xi}G_{4}, G_{4},  G^{n-1}\Big)+
(n-1)Str\Big( G^2_{4}, \delta_{\xi}G,  G^{n-2}\Big)\nn \\
&=&
2Str\Big([G_{4}, \xi], G_{4},  G^{n-1}\Big)+
(n-1)Str\Big( G^2_{4}, [G, \xi],  G^{n-2}\Big)=0,\nn
\eeqa
where in the last two equations we again used (B.10) of \cite{LBL-16443}.
However only the sum of these terms is a closed form.
We can check the closeness of the form $\Xi_{2n+6}$ by taking the exterior derivative:
\beqa
d\Xi_{2n+6}&=& Str\Big(DG_{6},  G^{n}\Big)+
 nStr\Big(G_{6}, DG,  G^{n-1}\Big) + \nn\\
&+&2nStr\Big(DG_{4}, G_{4},  G^{n-1}\Big)+
 n(n-1)Str\Big( G^2_{4}, DG, G^{n-2}\Big) =\nn \\
&=&2Str\Big([G_{4}, A_{3}],  G^{n}\Big)+Str\Big([G, A_{5}],  G^{n}\Big)+
2n Str\Big([G, A_{3}], G_{4},  G^{n-1}\Big)= \\
&=&2\bigg\{
Str\Big([G_{4}, A_{3}],  G^{n}\Big)+
nStr\Big(G_{4}, [G, A_{3}],  G^{n-1}\Big)\bigg\}+
  Str\Big([G, A_{5}],  G^{n}\Big)=0.\nn
\eeqa
On the first step we used (B.13) of \cite{LBL-16443} and on the second - the
Bianchi identities $DG=0$.
On the last step, the terms in the big brace as well as the last term are zero because
of (B.10) of \cite{LBL-16443}.
Again, according to Poincar\'e's lemma, this equation implies that $\Xi_{2n+6}$ can be
locally written as an exterior derivative of a certain (2n +5)-form.
In order to find that form we need to differentiate
$\Xi_{2n+6}$ over the deformation parameter t.  We have:
\beqa
&&{d\over dt}\Xi_{2n+6}(A_{t}, A_{3t}, A_{5t})=  Str\Big({dG_{6t}\over dt},  G_{t}^{n}\Big)+
nStr\Big(G_{6t}, {dG_{t}\over dt},  G_{t}^{n-1}\Big)+
2nStr\Big({dG_{4t}\over dt}, G_{4t},  G_{t}^{n-1}\Big)+\nn \\
&&+n(n-1)Str\Big( G^2_{4t}, {dG_{t}\over dt},  G_{t}^{n-2}\big)=
Str\Big(D_{t}A_{5},  G_{t}^{n}\Big)+
Str\Big(\{A, A_{5t}\},  G_{t}^{n}\Big)+
2Str\Big(\{A_{3}, A_{3t}\},  G_{t}^{n}\Big)+\nn \\
&&+nStr\Big(G_{6t}, D_{t}A,  G_{t}^{n-1}\Big)+
2nStr\Big(D_{t}A_{3}, G_{4t},  G_{t}^{n-1}\Big)+
2nStr\Big(\{A,A_{3t}\}, G_{4t},  G_{t}^{n-1}\Big)+\nn \\
&&+n(n-1)Str\Big( G^2_{4t}, D_{t}A,  G_{t}^{n-2}\Big)=
 dStr\Big(A_{5},  G_{t}^{n}\Big)+
Str\Big(\{A, A_{5t}\},  G_{t}^{n}\Big)+
2Str\Big(\{A, A_{3t}\},  G_{t}^{n}\Big)+\nn \\
&&+ndStr\Big(G_{6t}, A, G_{t}^{n-1}\Big)
-nStr\Big(D_{t}G_{6t}, A,  G_{t}^{n-1}\Big)+\nn \\
&&+2ndStr\Big(A_{3}, G_{4t}, G_{t}^{n-1}\Big)+
2nStr\Big(A_{3}, D_{t}G_{4t},  G_{t}^{n-1}\Big)+
2nStr\Big(\{A, A_{3t}\}, G_{4t},  G_{t}^{n-1}\Big)\nn \\
&&n(n-1)dStr\Big( G^2_{4t}, A,  G_{t}^{n-2}\Big)-
2n(n-1)Str\Big(D_{t}G_{4t}, G_{4t}, A,  G_{t}^{n-2}\Big)=\nn \\ \nn
\eeqa
\beqa
&= d\bigg\{
nStr\Big(G_{6t}, A,  G_{t}^{n-1}\Big)+
n(n-1)Str\Big( G^2_{4t}, A,  G_{t}^{n-2}\Big)+
2nStr\Big(G_{4t}, A_{3},  G_{t}^{n-1}\Big)
+Str\Big(A_{5},  G_{t}^{n}\Big)\bigg\}+\nn \\
&+ 2\bigg[Str\Big(\{A_{3}, A_{3t}\},  G_{t}^{n}\Big)
+nStr\Big(A_{3}, [G_{t}, A_{3t}],  G_{t}^{n-1}\bigg]+\nn \\
&+2n\bigg[Str\Big(\{A, A_{3t}\}, G_{4t},  G_{t}^{n-1}\Big)-
Str\Big([G_{4t}, A_{3t}], A,  G_{t}^{n-1}\bigg]+ \nn\\
&+2n\bigg[Str\Big(\{A, A_{3t}\}, G_{4t},  G_{t}^{n-1}\Big)+
Str\Big(A, [G_{4t}, A_{3t}],  G_{t}^{n-1}\Big)-
(n-1)Str\Big([G_{t}, A_{3t}], G_{4t}, A,  G_{t}^{n-2}\Big)
\bigg] +\nn\\
&+ Str\Big(\{A, A_{5t}\},  G_{t}^{n}\Big)-
nStr\Big([G_{t}, A_{5t}], A,  G_{t}^{n-1}\Big)=d \phi_{2n+5} .
\eeqa
On the second, third and last steps we used the relations (\ref{derivatives of zumino tensors})
and (B.13), (B.10) of \cite{LBL-16443} respectively.
Integrating the above equation over the parameter t in the interval $[0,1]$ we
shall get the following integral representation of the secondary form:
\beqa\label{secondaryform2}
\phi_{2n+5}&=&
\int_{0}^{1}dt\; \bigg\{
nStr\Big(G_{6t}, A,  G_{t}^{n-1}\Big)+
n(n-1)Str\Big(G^2_{4t}, A,   G_{t}^{n-2}\Big)+ \\
&&\qquad\qquad\qquad\qquad
+2nStr\Big( G_{4t}, A_{3}, G_{t}^{n-1}\Big)
+Str\Big(A_{5},  G_{t}^{n}\Big)\bigg\}.\nn
\eeqa
As we already mentioned above, the secondary form is not unique and it can be modified by the
addition of the differential of a one-step-lower-order form $d\beta_{2n+4}$ (\ref{cohomology}).
As in the case of $\psi_{2n+3}$, we will use this freedom in order to simplify our result.
Adding $d \beta_{2n+4}$, where
\[
\beta_{2n+4} = -\int_{0}^{1}dt\;
\bigg[
nStr \Big(A_{5t}, A,  G_{t}^{n-1}\Big)+
n(n-1)Str\Big(G_{4t}, A_{3t}, A,  G_{t}^{n-2}\Big)
\bigg],
\]
we get:
\beqa
\phi_{2n+5} &\backsim& \phi_{2n+5} +d \beta_{2n+4}=\nn \\
&=& \phi_{2n+5}+\int_{0}^{1}dt\; \bigg[
-nStr\Big(D_{t}A_{5t}, A,  G_{t}^{n-1}\Big)
+nStr\Big(A_{5t}, D_{t}A,  G_{t}^{n-1}\Big)-\nn \\
&&
-n(n-1) Str\Big(A_{5t}, A, D_{t}G_{t}, G_{t}^{n-2}\Big) -
n(n-1) Str\Big(D_{t}G_{4t}, A_{3t}, A,  G_{t}^{n-2}\Big)-\nn \\
&&
-n(n-1) Str\Big(G_{4t}, D_{t}A_{3t}, A, G_{t}^{n-2}\Big)+
n(n-1) Str\Big(G_{4t}, A_{3t}, D_{t}A,  G_{t}^{n-2}\Big)-\nn \\
&&
-n(n-1)(n-2) Str\Big(G_{4t}, A_{3t}, A, D_{t}G_{t},
 G_{t}^{n-3}\Big) \bigg],\nn
\eeqa
where one should use the Bianchi identities   $D_{t}G_{t}=0$
and (B.13) of \cite{LBL-16443}.
Next, with the aid of (\ref{field-strength tensors}) and (\ref{derivatives of zumino tensors})
we get,
\beqa
&&\phi_{2n+5} \backsim \phi_{2n+5}+ \int_{0}^{1}dt\; \bigg[
-nStr\Big(G_{6t}, A,  G_{t}^{n-1}\Big)+
nStr\Big(\{A_{3t}, A_{3t}\}, A,  G_{t}^{n-1}\Big)+\nn \\
&&\qquad
+nStr\Big(A_{5t}, {\partial G_{t}\over \partial t},  G_{t}^{n-1}\Big)-
n(n-1) Str\Big([G_{t}, A_{3t}], A_{3t}, A,  G_{t}^{n-2}\Big)-\nn \\
&&\qquad
-n(n-1)Str\Big( G^2_{4t}, A,  G_{t}^{n-2}\Big)+
n(n-1)Str\Big(G_{4t}, A_{3t}, {\partial G_{t}\over \partial t},
 G_{t}^{n-2}\Big)\bigg].\nn
\eeqa
One can see that
the first, the third and fifth terms cancel with the first two terms of $\phi_{2n+5}$
(\ref{secondaryform2}).
With the aid of (B.9) of \cite{LBL-16443}, the
second and the forth terms combine to give
$Str\Big(A_{3}, \{A,A_{3t}\},  G_{t}^{n-1}\Big)$.
Finally using the equation  $tD_{t}A_{3}=G_{4t}$ and (\ref{field-strength tensors})
we shall get the following expression for $\phi_{2n+5}$:
\beqa
&&
\int_{0}^{1}dt\; \bigg[
2nStr\Big(G_{4t}, A_{3},  G_{t}^{n-1}\Big)
+Str\Big(A_{5},  G_{t}^{n}\Big)+
nStr\Big(\{A, A_{3t}\}, A_{3t},  G_{t}^{n-1}\Big)+\nn \\
&&\qquad\qquad
+nStr\Big(A_{5t}, {\partial G_{t}\over \partial t}, G_{t}^{n-1}\Big)+
n(n-1)Str\Big(G_{4t}, A_{3t}, {\partial G_{t}\over \partial t},  G_{t}^{n-2}\Big)
\bigg]=\nn \\
&=&
\int_{0}^{1}dt\; \bigg[
Str\Big(A_{5},  G_{t}^{n}\Big)+
nStr\Big(A_{5t}, {\partial G_{t}\over \partial t},  G_{t}^{n-1}\Big)+\nn \\
&&\qquad\qquad
+nStr\Big(D_{t}A_{3}+t\{A,A_{3}\}, A_{3t},  G_{t}^{n-1}\Big)+
nStr\Big(G_{4t}, A_{3},  G_{t}^{n-1}\Big)+\nn \\
&&\qquad\qquad
+n(n-1)Str\Big(G_{4t}, A_{3t}, {\partial G_{t}\over \partial t},  G_{t}^{n-2}\Big)\bigg]=\nn \\
&=&
\int_{0}^{1}dt\;
{\partial\over \partial t}\bigg[
Str\bigg(
A_{5t},  G_{t}^{n}\bigg)
+nStr\bigg(
G_{4t}, A_{3t},  G_{t}^{n-1}\bigg)\bigg].\nn
\eeqa
Hence, after the integration we get
\beqa\label{secondary2}
\phi_{2n+5}=Str\bigg(
A_{5},  G^{n}\bigg)
+nStr\bigg(
A_{3}, G_{4},  G^{n-1}\bigg).
\eeqa
By comparing the representations of the secondary form $\phi_{2n+5}$ in (\ref{secondaryform2})
and in (\ref{secondary2}) it becomes clear  that the last expression is much more simple and transparent.
Let us verify that the exterior derivative of the above form leads us back to $\Xi_{2n+6}$.
\beqa
&&d \phi_{2n+5}=Str\bigg(DA_{5},  G^{n}\bigg)-
n  Str\bigg(A_{5}, DG,  G^{n-1}\bigg) +
nStr\bigg(DG_{4}, A_{3},  G^{n-1}\bigg)+\nn \\
+&&nStr\bigg(G_{4}, DA_{3},  G^{n-1}\bigg)
-n(n-1) Str\bigg(
G_{4}, A_{3}, DG,  G^{n-2}\bigg) =\nn \\
=&& Str\bigg( G_{6},  G^{n}\bigg)-
Str\bigg( \{A_{3}, A_{3}\},  G^{n}\bigg)+ nStr\bigg([G, A_{3}], A_{3},  G^{n-1}\bigg)+\nn \\
&&+ n Str\bigg( G^2_{4},  G^{n-1}\bigg)=
 Str\bigg(G_{6},  G^{n}\bigg)
+nStr\bigg( G^2_{4},  G^{n-1}\bigg)
=\Xi_{2n+6},\nn
\eeqa
where on the second step the second and third terms cancel because of (B.10) of \cite{LBL-16443}.

In order to find out the potential anomalies we have to calculate the gauge variation of the secondary
form $\phi_{2n+5}$ with respect to the scalar, rank-2 and rank-4 gauge parameters.
We have
\beqa
\delta_{\xi}\phi_{2n+5}&=&
\delta_{\xi}\bigg[Str\Big(A_{5}, G^{n}\Big)
+nStr\Big(G_{4}, A_{3},  G^{n-1}\Big)\bigg]=\nn \\
&=&
Str\Big(\delta_{\xi}A_{5},  G^{n}\bigg)+
nStr\Big(A_{5}, \delta_{\xi}G,  G^{n-1}\bigg)+
nStr\Big(\delta_{\xi}G_{4}, A_{3},  G^{n-1}\Big)+\nn \\
&&+
nStr\Big(G_{4}, \delta_{\xi}A_{3},  G^{n-1}\Big)+
n(n-1)Str\Big(G_{4}, A_{3}, \delta_{\xi}G,  G^{n-2}\Big)=\nn \\
&=&
Str\Big([A_{5},\xi],  G^{n}\bigg)+
nStr\Big(A_{5}, [G, \xi],  G^{n-1}\bigg)+
nStr\Big([G_{4}, \xi], A_{3},  G^{n-1}\Big)+\nn \\
&&+
nStr\Big(G_{4}, [A_{3}, \xi], G^{n-1}\Big)+
n(n-1)Str\Big(G_{4}, A_{3}, [G, \xi],  G^{n-2}\Big)=0.\nn
\eeqa
Thus, there are no anomalies in the standard gauge symmetry. But there are potential anomalies in
the higher-rank gauge symmetries. Indeed,
\beqa
\delta_{\zeta_{2}}\phi_{2n+5}&=&Str\Big(\delta_{\zeta_{2}}A_{5},  G^{n}\Big)+
 nStr\Big(\delta_{\zeta_{2}}G_{4}, A_{3},  G^{n-1}\Big)+
nStr\Big(G_{4}, \delta_{\zeta_{2}}A_{3},  G^{n-1}\Big) =\nn \\
&=&2Str\Big([A_{3}, \zeta_{2}],  G^{n}\Big)+
nStr\Big([G, \zeta_{2}], A_{3},  G^{n-1}\Big)+
nStr\Big(G_{4}, D\zeta_{2}, G^{n-1}\Big)=\nn \\
&=&Str\Big([A_{3}, \zeta_{2}],  G^{n}\Big)+
nStr\Big(G_{4}, D\zeta_{2},  G^{n-1}\Big)=\nn \\
&=&Str\Big([A_{3}, \zeta_{2}],  G^{n}\Big)+
ndStr\Big(G_{4}, \zeta_{2},  G^{n-1}\Big)-
nStr\Big([G, A_{3}], \zeta_{2}, G^{n-1}\Big)=\nn \\
&=& ndStr\Big(\zeta_{2}, G_{4},  G^{n-1}\Big)
\eeqa
and
\beqa
\delta_{\zeta_{4}}\phi_{2n+5}=Str\Big(D\zeta_{4},  G^{n}\Big)=
dStr\Big(\zeta_{4},  G^{n}\Big).
\eeqa
Hence the anomalies are:
\beqa\label{amomaly2}
\phi^{(1)}_{2n+4}(\zeta_{4},A)&=&Str\Big(\zeta_{4},  G^{n}\Big)\nn\\
\phi^{(1)}_{2n+4}(\zeta_{2},A, A_{3})&=&nStr\Big(\zeta_{2}, G_{4},  G^{n-1}\Big)
\eeqa
{\it In summary, we have the expressions (\ref{tensoranomaly2}) for primary form $\Xi_{2n+6}$,
the expression (\ref{secondary2}) for the secondary form $\phi_{2n+5}$ and
(\ref{amomaly2}) for the anomalies.}

\newpage

\section{\it The Form $\Upsilon_{2n+8}$}

In the recent article \cite{Antoniadis:2013jja} the authors found the following exact,
metric independent forms, linear in $G_{8}$:
\beqa\label{Y10}
\Upsilon_{10}&=&Tr (GG_{8}+3G_{4}G_{6})=
Str(G_{8}, G)+3Str(G_{4}, G_{6}),\nn\\
\Upsilon_{12}&=&Tr (G^{2}G_{8}+3GG_{4}G_{6}+3GG_{6}G_{4}+2G_{4}^{3})=\nn\\
&=&Str(G_{8}, G^{2})+6Str(G_{4}, G_{6}, G)+2Str(G_{4}^{3}).
\eeqa
In order to find the general expression for the forms linear in $G_8$ let us first find the next form
$\Upsilon_{14}$. For that let us consider the linear combination of all possible rank-14 Lorentz
invariant traces which can be constructed in terms of field strength tensors:
\beqa
\Upsilon_{14}&=&Tr\bigg(
G^{3}G_{8}+c_{1}GG_{6}^{2}+c_{2}G_{4}^{2}G_{6}+c_{3}G^{2}G_{4}G_{6}+
c_{4}GG_{4}GG_{6}+c_{5}G^{2}G_{6}G_{4}+
\nonumber\\
&&\qquad
+c_{6}G^{4}G_{6}+c_{7}GG_{4}^{3}+c_{8}G^{3}G_{4}^{2}+
c_{9}GG_{4}GG_{4}G +c_{10}G^{5}G_{4}+c_{11}G^{7} \bigg).
\eeqa
The two terms with the coefficients $c_{6}=c_{8}$ can be dropped  since
they compose the form $\Xi_{14}$ (\ref{tensoranomaly2}). The terms with coefficients
$c_{10}$ and $c_{11}$ can also be dropped since they represent the forms $\Phi_{14}$
(\ref{tensoranomaly}) and $\CP_{14}$  (\ref{abeliananom}) respectively.
Using the Bianchi identities (\ref{Bianchi Identities})
one can calculate the derivatives of the following terms:
\beqa
d ~Tr(G^{3}G_{8})=3 Tr (G^{3}(G_{6} A_{3}- A_{3} G_{6}  + G_{4} A_{5}-A_{5} G_{4}  )),\nn\\
d~Tr( G^{2}G_{4}G_{6})=Tr ( G^{3}(A_{3}G_{6}-G_{4}A_{5})
+2G^{2}(G_{4}^{2}A_{3}-G_{4}A_{3}G_{4})+
G^{2}G_{4}GA_{5}-G^{2}A_{3}GG_{6} ),\nn\\
d~Tr(GG_{4}GG_{6})=Tr (GA_{3}(GG_{6}G-G^{2}G_{6})+2GG_{4}G(G_{4}A_{3}-A_{3}G_{4})+
(GG_{4}G^{2}-G^{2}G_{4}G)A_{5}),\nn\\
d~Tr(G^{2}G_{6}G_{4})=Tr (2G^{2}(G_{4}A_{3}G_{4}-A_{3}G_{4}^{2})+G^{2}(G_{6}GA_{3}-GG_{6}A_{3})+
G^{2}(GA_{5}-A_{5}G)G_{4}),\nn\\
d~Tr(GG_{4}^{3})=Tr (G^{2}(A_{3}G_{4}^{2}-G_{4}^{2}A_{3})+GG_{4}G(A_{3}G_{4}-G_{4}A_{3})),\nn
\eeqa
and see that the following combination is a closed form:
\beqa
d~Tr(G^{3}G_{8}+3G^{2}G_{4}G_{6}+3GG_{4}GG_{6}+3G^{2}G_{6}G_{4}+6GG_{4}^{3})=0.
\eeqa
Hence,
\beqa \label{Y14}
\Upsilon_{14}&=&Tr (G^{3}G_{8}+3G^{2}G_{4}G_{6}+3G^{2}G_{6}G_{4}+3GG_{4}GG_{6}+6GG_{4}^{3})=\nn\\
&=&Str(G_{8}, G^{3}) +9Str(G_{4}, G_{6}, G^2)+6Str(G, G_{4}^{3})
\eeqa
and it can be written in terms of symmetric trace.  The rest of the terms with the coefficients $c_{1},c_{2},c_{9}$ do not comprise any closed
form.

One can check that the  $\Upsilon_{14}$ is gauge invariant.
In terms of the standard gauge parameter we get:
\beqa
\delta_{\xi} \Upsilon_{14}&=&   Tr \bigg[ [G^{3},\xi]G_{8}+G^{3}\Big( [G_{8},\xi]+3[G_{6},\zeta_{2}]+3[G_{4},\zeta_{4}]+[G,\zeta_{6}]\Big)+
3[G^{2},\xi]G_{4}G_{6}+
\nonumber \\
&&+3G^{2}\Big([G_{4},\xi]+[G,\zeta_{2}]\Big)G_{6}+3G^{2}G_{4}\Big([G_{6},\xi]+2[G_{4},\zeta_{2}]+[G,\zeta_{4}]\Big)+
\nonumber\\
&&+3[G^{2},\xi]G_{6}G_{4}+3G^{2}\Big( [G_{6},\xi]+2[G_{4},\zeta_{2}]+[G,\zeta_{4}]\Big)G_{4}+
3G^{2}G_{6}\Big( [G_{4},\xi]+[G,\zeta_{2}]\Big)+
\nonumber\\
&&+3[G,\xi]G_{4}GG_{6}+3G\Big( [G_{4},\xi]+[G,\zeta_{2}]\Big) GG_{6}+3GG_{4}[G,\xi]G_{6}+
\nonumber\\
&&+3GG_{4}G\Big( [G_{6},\xi]+2[G_{4},\zeta_{2}]+[G,\zeta_{4}]\Big)
+6[G,\xi]G_{4}^{3}+6G  [G_{4}^{3},\xi]
\bigg]=0.
\eeqa
In an analogous way one can easily prove that $\Upsilon_{14}$ is invariant under the transformations
of the higher-tensor gauge parameters $\zeta_{2}$, $\zeta_{4}$, $\zeta_{6}$.

Having in hand the series of forms $\Upsilon_{10,12,14}$ one can guess a general expression
for $\Upsilon_{2n+8}$ and check that it fulfills all the required properties.
We suggest the following general form for $\Upsilon_{2n+8}$:
\beqa \label {Y2n+8}
\Upsilon_{2n+8}=Str\bigg(G_{8}, G^{n}\bigg)+3n Str\bigg(G_{4}, G_{6},
 G^{n-1}\bigg)+
n(n-1) Str\bigg(G^3_{4}, G^{n-2}
\bigg).
\eeqa
As one can see, each term of the $\Upsilon_{2n+8}$ is separately gauge invariant.
Variating over the standard gauge parameter we get:
\beqa
\delta_{\xi}Str\Big(G_{8}, G^{n}\Big)&=&Str\Big(\delta_{\xi}G_{8}, G^{n}\Big)
+nStr\Big(G_{6}, \delta G,  G^{n-1}\Big)=\nn \\
&=&Str\Big([G_{8}, \xi], G^{n}\Big)+
nStr\Big(G_{8}, [G, \xi],  G^{n-1}\Big)=0, \nn\\
\delta_{\xi}Str\Big(G_{4}, G_{6},  G^{n-1}\Big)&=&
Str\Big([G_{4}, \xi], G_{6},  G^{n-1}\Big)+
Str\Big(G_{4}, [G_{6}, \xi],  G^{n-1}\Big)+\nn \\
&&\quad +
(n-1)Str\Big(G_{4}, G_{6}, [G,\xi], ] G^{n-2}\Big)=0,\nn\\
\delta_{\xi}Str\Big( G^3_{4},  G^{n-1}\Big)&=&
3Str\Big(\delta_{\xi}G_{4},  G^2_{4},
 G^{n-1}\Big)+
(n-1)Str\Big( G^3_{4}, \delta_{\xi}G, G^{n-2}\Big) =\nn \\
&=&
3Str\Big([G_{4}, \xi],  G^2_{4},G^{n-1}\Big)+
(n-1)Str\Big( G^3_{4}, [G,\xi],
 G^{n-2}\Big) =0
\eeqa
Analogously, one can check that each term of $\Upsilon_{2n+8}$ is separately
invariant under the variation over the higher-tensor gauge parameters.
The last two calculations clearly demonstrate the power of the use of the symmetrized
traces and of their properties.

Taking the exterior derivative of $\Upsilon_{2n+8}$  one can become convinced that it is a closed  form:
\beqa
d\Upsilon_{2n+8}
&=& Str\Big(DG_{8}, G^{n}\Big)+
nStr\Big(G_{8}, DG,  G^{n-1}\Big)+
3nStr\Big(DG_{4}, G_{6},  G^{n-1}\Big)+\nn \\
&&+
3nStr\Big(G_{4}, DG_{6},  G^{n-1}\Big)+
3n(n-1)Str\Big(G_{4}, G_{6}, DG, G^{n-2}\Big)+\nn \\
&&+
3n(n-1)Str\Big(DG_{4},  G^2_{4}, G^{n-2}\Big)+
n(n-1)(n-2)Str\Big( G^3_{4}, DG,  G^{n-3}\Big)\nn\\
&=&
3Str\Big([G_{6}, A_{3}], G^{n}]\Big)+3 Str\Big([G_{4}, A_{5}], G^{n}\Big)+
Str\Big([G,A_{7}], G^{n}\Big)+ \nn \\
&&+ 3n Str\Big([G,A_{3}], G_{6},  G^{n-1}\Big)+
6n Str\Big(G_{4}, [G_{4},A_{3}],  G^{n-1}\Big)+\nn \\
&&+
3n Str\Big(G_{4}, [G,A_{5}],  G^{n-1}\Big)+
3n(n-1)Str\Big([G,A_{3}], G_{4}, G_{4},  G^{n-2}\Big)=\nn \\
&=&
3\bigg\{
Str\Big([G_{6}, A_{3}], G^{n}]\Big)+
n Str\Big(G_{6}, [G,A_{3}],  G^{n-1}\Big) \bigg\}+ \nn\\
&&+
3n\bigg\{
2Str\Big([G_{4},A_{3}], G_{4} ,  G^{n-1}\Big)+
(n-1)Str\Big(G_{4}, G_{4}, [G,A_{3}],  G^{n-2}\Big)\bigg\}+\nn \\
&&+
3\bigg\{
Str\Big([G_{4}, A_{5}],  G^{n}\Big)+
n Str\Big(G_{4}, [G,A_{5}],  G^{n-1}\Big) \bigg\}+\nn \\
&&+
{1\over n+1}(n+1)Str\Big([G,A_{7}],  G^{n}\Big)=0.
\eeqa
Again, according to Poincar\'e's lemma, this equation implies that $\Upsilon_{2n+8}$ can be
locally written as an exterior derivative of a certain (2n+7)-form $\rho_{2n+7}$.
In order to find that form we need to differentiate $\Upsilon_{2n+8}$ over the deformation parameter $t$,
as we did in the previous section:
\beqa
&&{d\over dt}\Upsilon_{2n+8} =
Str\Big({dG_{8t}\over dt},  G_{t}^{n}\Big)+
nStr\Big(G_{8t}, {dG_{t}\over dt},  G_{t}^{n-1}\Big)+
3nStr\Big({dG_{4t}\over dt}, G_{6t},  G_{t}^{n-1}\Big)+\nn \\
&&+3nStr\Big(G_{4t}, {dG_{6t}\over dt},  G_{t}^{n-1}\Big)+
3n(n-1)Str\Big(G_{4t}, G_{6t}, {dG_{t}\over dt},  G_{t}^{n-2}\Big)+\nn \\
&&+3n(n-1)Str\Big({dG_{4t}\over dt},  G^2_{4t},  G_{t}^{n-2}\Big)+
n(n-1)(n-2)Str\Big(G_{4t}, G^2_{4t}, {dG_{t}\over dt},  G_{t}^{n-3}\Big)\nn \\
=&& Str\Big(D_{t}A_{7},  G_{t}^{n}\Big)+
6Str\Big(\{A_{3}, A_{5t}\},  G_{t}^{n}\Big)+
Str\Big(\{A, A_{7t}\},  G_{t}^{n}\Big)+\nn \\
&&+nStr\Big(G_{8t}, D_{t}A,  G_{t}^{n-1}\Big)+
3nStr\Big(D_{t}A_{3}, G_{6t},  G_{t}^{n-1}\Big)+
+3nStr\Big(\{A, A_{3t}\}, G_{6t},  G_{t}^{n-1}\Big)+\nn \\
&&+3nStr\Big(G_{4t}, D_{t}A_{5},  G_{t}^{n-1}\Big)+
3nStr\Big(G_{4t}, \{A, A_{5t}\}, G_{t}^{n-1}\Big)+
6nStr\Big(G_{4t}, \{A_{3}, A_{3t}\}, G_{t}^{n-1}\Big)\nn \\
&&+3n(n-1)Str\Big(G_{4t}, G_{6t}, D_{t}A,  G_{t}^{n-2}\Big)+
3n(n-1)Str\Big(D_{t}A_{3}, G^2_{4t},  G_{t}^{n-2}\Big)+\nn \\
&&+3n(n-1)Str\Big(\{A, A_{3t}\}, G^2_{4t},  G_{t}^{n-2}\Big)+
n(n-1)(n-2)Str\Big( G^3_{4t}, D_{t}A,  G_{t}^{n-3}\Big).\nn
\eeqa
In some of the terms we can extract the covariant exterior derivatives
outside the symmetrized traces:
\beqa
&&{d\over dt}\Upsilon_{2n+8}=dStr\Big(A_{7},  G_{t}^{n}\Big)+
6Str\Big(\{A_{3}, A_{5t}\},  G_{t}^{n}\Big)+
Str\Big(\{A, A_{7t}\},  G_{t}^{n}\Big)+\nn \\
&&+ndStr\Big(G_{8t}, A,  G_{t}^{n-1}\Big)-
nStr\Big(D_{t}G_{8t}, A,  G_{t}^{n-1}\Big)+
3ndStr\Big(A_{3}, G_{6t},  G_{t}^{n-1}\Big)+\nn \\
&&+3nStr\Big(A_{3}, D_{t}G_{6t},  G_{t}^{n-1}\Big)+
3nStr\Big(\{A, A_{3t}\}, G_{6t}, G_{t}^{n-1}\Big)+
3ndStr\Big(G_{4t}, A_{5}, G_{t}^{n-1}\Big)-\nn \\
&&-3nStr\Big(D_{t}G_{4t}, A_{5},  G_{t}^{n-1}\Big)+
3nStr\Big(G_{4t}, \{A, A_{5t}\},  G_{t}^{n-1}\Big)+
6nStr\Big(G_{4t}, \{A_{3}, A_{3t}\},  G_{t}^{n-1}\Big)+\nn\\
&&+3n(n-1)dStr\Big(G_{4t}, G_{6t}, A,  G_{t}^{n-2}\Big)-
3n(n-1)Str\Big(D_{t}G_{4t}, G_{6t}, A,  G_{t}^{n-2}\Big)-\nn \\
&&-3n(n-1)Str\Big(G_{4t}, D_{t}G_{6t}, A, G_{t}^{n-2}\Big)+
3n(n-1)dStr\Big(A_{3}, G_{4t}, G_{4t},  G_{t}^{n-2}\Big)+\nn \\
&&+6n(n-1)Str\Big(A_{3}, D_{t}G_{4t}, G_{4t},  G_{t}^{n-2}\Big)+
3n(n-1)Str\Big(\{A, A_{3t}\}, G_{4t}, G_{4t}, G_{t}^{n-2}\Big)+\nn \\
&&+n(n-1)(n-2)dStr\Big(G^3_{4t}, A,  G_{t}^{n-3}\Big)
-3n(n-1)(n-2)Str\Big(D_{t}G_{4t},  G^2_{4t}, A,  G_{t}^{n-3}\Big).\nn
\eeqa
As one can see, some of the terms are written as exterior derivatives. We shall collect
them in the formula below and then combine the rest of the terms in square brackets:
\beqa
&{d\over dt}\Upsilon_{2n+8}= d\bigg\{
Str\Big(A_{7},  G_{t}^{n}\Big)+
nStr\Big(G_{8t}, A,  G_{t}^{n-1}\Big)+
3nStr\Big(A_{3}, G_{6t}, G_{t}^{n-1}\Big)+\nn \\
& \qquad+3n\Big(G_{4t}, A_{5},  G_{t}^{n-1}\Big)+
3n(n-1)Str\Big(G_{4t}, G_{6t}, A, G_{t}^{n-2}\Big)+\nn \\
& \qquad+3n(n-1)Str\Big(A_{3},  G^2_{4t},  G_{t}^{n-2}\Big)+
n(n-1)(n-2)Str\Big( G^3_{4t}, A,  G_{t}^{n-3}\Big)\bigg\}+\nn \\
& +3\bigg[
Str\Big(\{A_{3}, A_{5t}\},  G_{t}^{n}\Big)+
nStr\Big(A_{3}, [G_{t}, A_{5t}],  G_{t}^{n-1}\Big)\bigg]+\nn \\
& +3\bigg[
Str\Big(\{A_{3t}, A_{5}\}, G_{t}^{n}\Big)-
nStr\Big([G_{t}, A_{3t}], A_{5},  G_{t}^{n-1}\Big)\bigg]+\nn \\
& +Str\Big(\{A, A_{7t}\},  G_{t}^{n}\Big)-
nStr\Big([G_{t}, A_{7t}], A,  G_{t}^{n-1}\Big)+\nn \\
& +3n\bigg[
Str\Big(\{A, A_{3t}\}, G_{6t},  G_{t}^{n-1}\Big)-
Str\Big([G_{6t}, A_{3t}], A,  G_{t}^{n-1}\Big)-
(n-1)Str\Big([G_{4}, A_{3t}], G_{6t}, A,  G_{t}^{n-1}\Big)
\bigg] \nn \\
& +3n\bigg[
Str\Big(G_{4t}, \{A, A_{5t}\},  G_{t}^{n-1}\Big)-
Str\Big([G_{4t}, A_{5t}], A,  G_{t}^{n-1}\Big)
-(n-1)Str\Big(G_{4t}, [G_{t}, A_{5t}], A,  G_{t}^{n-2}\Big)\bigg]\nn \\
& +6n\bigg[
Str\Big(A_{3}, [G_{4t}, A_{3t}],  G_{t}^{n-1}\Big)+
Str\Big(G_{4t}, \{A_{3}, A_{3t}\}, G_{t}^{n-1}\Big)
+(n-1)Str\Big(A_{3}, [G_{t}, A_{3t}], G_{4t},  G_{t}^{n-2}\Big)\bigg]\nn \\
& +3n(n-1)\bigg[
Str\Big(\{A, A_{3t}\}, G_{4t}, G_{4t},  G_{t}^{n-2}\Big)-
2Str\Big(G_{4t}, [G_{4t}, A_{3t}], A,  G_{t}^{n-2}\Big)-\nn \\
& \qquad\qquad\qquad\qquad\qquad\qquad
-(n-2)Str\Big([G_{t}, A_{3t}],  G^2_{4t}, A, G_{t}^{n-3}\Big)
\bigg].\nn
\eeqa
The terms in the square brackets vanish thanks to the identity (B.10).
Therefore we have the following integral representation for the secondary form:
\beqa\label{seccondary3}
\rho_{2n+7}&=&\int_{0}^{1}dt\bigg\{
n Str\Big(G_{8t}, A,  G_{t}^{n-1}\Big)+
3n(n-1)Str\Big(G_{4t}, G_{6t}, A,  G_{t}^{n-2}\Big)+\nn \\
&& +
n(n-1)(n-2)Str\Big( G^3_{4t}, A,  G_{t}^{n-3}\Big)+
3nStr\Big(G_{6t}, A_{3},  G_{t}^{n-1}\Big)+\nn \\
&& +
3n(n-1)Str\Big( G^2_{4t}, A_{3},  G_{t}^{n-2}\Big)+
3nStr\Big(G_{4t}, A_{5},  G_{t}^{n-1}\Big)+
Str\Big(A_{7},  G_{t}^{n}\Big)\bigg\}.
\eeqa
As we already discussed in the introduction and in the previous section, the secondary forms
are defined modulo exact forms and in the given case up to (2n+7)-form
$\rho_{2n+7} \backsim \rho_{2n+7} + d \gamma_{2n+6}$. Therefore we have to choose an appropriate
candidate for $\gamma_{2n+6}$. It appears that to simplify the result the exterior derivative of the
following form should be subtracted:
\beqa
&&\gamma_{2n+6} =
\int_{0}^{1}dt\bigg[
nStr\Big(A_{7t}, A,  G_{t}^{n-1}\Big)+
n(n-1)(n-2)Str\Big( G^2_{4t}, A_{3t}, A,  G_{t}^{n-3}\Big)+\nn \\
&&
+n(n-1)Str\Big(G_{6t}, A_{3t}, A, G_{t}^{n-2}\Big)
+2n(n-1)Str\Big(G_{4t}, A_{5t}, A,  G_{t}^{n-2}\Big)
-nStr\Big(A_{3t}, A_{5},  G_{t}^{n-1}\Big)\bigg] .\nn
\eeqa
Subtracting the first two terms of the  $d \gamma_{2n+6}$  from $\rho_{2n+7}$ we will get:
\beqa
&&
Str\Big(A_{7},  G^{n}\Big)+
n(n-1)Str\Big( G^2_{4}, A_{3},  G^{n-2}\Big)+\nn \\
&&+\int_{0}^{1}\bigg\{
3n(n-1)Str\Big(G_{4t}, G_{6t}, A,  G_{t}^{n-2}\Big)+
3nStr\Big(G_{6t}, A_{3},  G_{t}^{n-1}\Big)+
3nStr\Big(G_{4t}, A_{5},  G_{t}^{n-1}\Big)+\nn \\
&&\qquad\qquad\qquad\qquad
+3nStr\Big(\{A_{3t}, A_{5t}\}, A,  G_{t}^{n-1}\Big)-
2n(n-1)Str\Big(\{A, A_{3t}\}, G_{4t}, A_{3t},  G_{t}^{n-2}\Big)-\nn \\
&&\qquad\qquad\qquad\qquad
-2n(n-1)(n-2)Str\Big([G_{t}, A_{3t}], G_{4t}, A_{3t}, A,  G_{t}^{n-3}\Big)\bigg\}.\nn
\eeqa
Subtracting now the last three terms of the $d \gamma_{2n+6}$, we will get:
\beqa\label{secondary3}
&\rho_{2n+7}=Str\Big(A_{7},  G_{t}^{n}\Big)+
n(n-1)Str\Big(G_{4}, G_{4}, A_{3}, G_{t}^{n-2}\Big)+ \\
&+
nStr\Big(G_{6t}, A_{3},  G_{t}^{n-1}\Big)+
2nStr\Big(G_{4}, A_{5},  G_{t}^{n-1}\Big).\nn
\eeqa
Let us check that the exterior derivative of the simplified secondary form
gives us back the primary form $\Upsilon_{2n+8}$. We have,
\beqa
&&d\rho_{2n+7}=Str \bigg(DA_{7}, G^{n}\bigg) +
nStr\bigg(DG_{6}, A_{3},  G^{n-1}\bigg) +nStr\bigg(G_{6}, DA_{3},  G^{n-1}\bigg)+\nn \\
&&+2nStr\bigg(DG_{4}, A_{5},  G^{n-1}\bigg)
+2nStr\bigg(G_{4}, DA_{5},  G^{n-1}\bigg)+\nn \\
&&+2n(n-1)Str\bigg(DG_{4}, G_{4}, A_{3},  G^{n-2}\bigg)
+n(n-1)Str\bigg( G^2_{4}, DA_{3},  G^{n-2}\bigg)=\nn\\
&=&Str\bigg(G_{8},  G^{n}\bigg)
-(2+1)Str\bigg(\{A_{3}, A_{5}\},  G^{n}\bigg)+
2nStr\bigg([G_{4}, A_{3}], A_{3},  G^{n-1}\bigg)+\nn \\
&&+nStr\bigg([G,A_{5}], A_{3},  G^{n-1}\bigg)-
nStr\bigg(G_{6}, G_{4},  G^{n-1}\bigg)+
2nStr\bigg([G, A_{3}], A_{5},  G^{n-1}\bigg)+\nn \\
&&+2nStr\bigg(G_{4}, G_{6}, G^{n-1}\bigg)-
2nStr\bigg(G_{4}, \{A_{3}, A_{3}\},  G^{n-1}\bigg)+\nn \\
&&+2n(n-1)Str\bigg([G,A_{3}], G_{4}, A_{3}, G^{n-2}\bigg)+
n(n-1)Str\bigg( G^3_{4},  G^{n-2}\bigg)
=\Upsilon_{2n+8}.\nn
\eeqa
Due to (B.10) of \cite{LBL-16443} the first part of the second term cancels with the sixth one,
the second part of the second term cancels with the forth one, and the third term cancels with
the ninth one.
The secondary form allows to find the potential  anomalies of the theory by performing the
transgression steps.
Thus in order to find out potential anomalies we have to calculate the gauge variation of the secondary
form $\rho_{2n+7}$ with respect to the scalar, rank-2, rank-4 and rank-6 gauge parameters:
\beqa
\delta_{\xi}\rho_{2n+7}&=&Str\Big([A_{7}, \xi],  G^{n}\Big)+
nStr\Big(A_{7}, [G, \xi],  G^{n-1}\Big)+\nn \\
&&+
nStr\Big([G_{6}, \xi], A_{3},  G^{n-1}\Big)+
nStr\Big(G_{6}, [A_{3}, \xi],  G^{n-1}\Big)+\nn \\
&&+
n(n-1)Str\Big(G_{6}, A_{3}, [G, \xi],  G^{n-2}\Big)+
2nStr\Big([G_{4}, \xi], A_{5},  G^{n-1}\Big)\nn \\
&&+
2nStr\Big(G_{4}, [A_{5}, \xi],  G^{n-1}\Big)+
2n(n-1)Str\Big(G_{4}, A_{5}, [G, \xi],  G^{n-2}\Big)+\nn \\
&&+
2n(n-1)Str\Big([G_{4}, \xi], G_{4}, A_{3},  G^{n-2}\Big)+
n(n-1)Str\Big( G^2_{4}, [A_{3}, \xi],  G^{n-2}\Big)+\nn \\
&& +
n(n-1)(n-2)Str\Big( G^2_{4}, A_{3}, [G, \xi], G^{n-3}\Big)=0,
\eeqa
where the identity (B.13) of \cite{LBL-16443} was used.
There are no anomalies in the standard gauge symmetry.

The variation over the rank-2 gauge
parameter gives:
\beqa
\delta_{\zeta_{2}}\rho_{2n+7}&=&3Str\Big([A_{5}, \zeta_{2}],  G^{n}\Big)+
2nStr\Big([G_{4}, \zeta_{2}], A_{3},  G^{n-1}\Big)+
nStr\Big(G_{6}, D\zeta_{2},  G^{n-1}\big)+\nn \\
&&+2nStr\Big([G, \zeta_{2}], A_{5}, G^{n-1}\Big)+
4nStr\Big(G_{4}, [A_{3}, \zeta_{2}],  G^{n-1}\Big)+\nn \\
&&+2n(n-1)Str\Big([G, \zeta_{2}], G_{4}, A_{3},  G^{n-2}\Big)+
n(n-1)Str\Big( G^2_{4}, D\zeta_{2},  G^{n-2}\Big)=\nn \\
&=&Str\Big([A_{5}, \zeta_{2}],  G^{n}\Big)+
ndStr\Big(G_{6}, \zeta_{2},  G^{n-1}\Big)-
nStr\Big(DG_{6}, \zeta_{2},  G^{n-1}\Big)+\nn \\
&&+2nStr\Big(G_{4}, [A_{3}, \zeta_{2}],  G^{n-1}\Big)+
n(n-1)dStr\Big( G^2_{4}, \zeta_{2},  G^{n-2}\Big)-\nn \\
&&-2n(n-1)Str\Big(DG_{4}, G_{4}, \zeta_{2},  G^{n-2}\Big)=\nn \\
&=&
d\bigg\{nStr\Big(\zeta_{2}, G_{6},  G^{n-1}\Big)+
n(n-1)Str\Big(\zeta_{2},  G^2_{4},  G^{n-2}\Big)\bigg\}+\nn \\
&&+Str\Big([A_{5}, \zeta_{2}],  G^{n}\Big)+
nStr\Big([A_{5}, G], \zeta_{2},  G^{n-1}\Big)+\nn \\
&&+2nStr\Big([A_{3}, G_{4}], \zeta_{2},  G^{n-1}\Big)+
2nStr\Big(G_{4}, [A_{3}, \zeta_{2}],  G^{n-1}\Big)-\nn \\
&& ~~~~
-2n(n-1)Str\Big([G, A_{3}], G_{4}, \zeta_{2},  G^{n-2}\Big),
\eeqa
where the sum of the third and the fourth terms   and  the sum of the last three
terms vanish due to (B.10). The variation over the rank-4 gauge parameter gives:
\beqa
\delta_{\zeta_{4}}\rho_{2n+7}&=&3Str\Big([A_{3}, \zeta_{4}], G^{n}\Big)+
nStr\Big([G, \zeta_{4}], A_{3},  G^{n-1}\Big)+
2nStr\Big(G_{4}, D\zeta_{4}, G^{n-1}\Big)=\nn \\
&=&
2Str\Big([A_{3}, \zeta_{4}],  G^{n}\Big)+
2ndStr\Big(G_{4}, \zeta_{4},  G^{n-1}\Big)-
2nStr\Big(DG_{4}, \zeta_{4}, G^{n-1}\Big)=\nn \\
&=&
2Str\Big([A_{3}, \zeta_{4}],  G^{n}\Big)+
2nStr\Big([A_{3}, G], \zeta_{4},  G^{n-1}\Big)+
2ndStr\Big(\zeta_{4}, G_{4},  G^{n-1}\Big)\nn \\
&=&2ndStr\Big(\zeta_{4}, G_{4},  G^{n-1}\Big)
\eeqa
and the variation over rank-6 gauge parameter is:
\beqa
\delta_{\zeta_{6}}\rho_{2n+7}=Str\Big(D\zeta_{6},  G^{n}\Big)=
dStr\Big(\zeta_{6},  G^{n}\Big).
\eeqa
Hence the corresponding anomalies are:
\beqa\label{amomaly3}
\rho_{2n+6}^{(1)} (\zeta_{6},A)&=&Str\Big(\zeta_{6}, G^{n}\Big),\nn\\
\rho^{(1)}_{2n+6}(\zeta_{4},A,A_{3})&=&
2n Str\Big(\zeta_{4}, G_{4},  G^{n-1}\Big),\nn\\
\rho^{(1)}_{2n+6}(\zeta_{2},A,A_{3},A_{5})&=&
n Str\Big(\zeta_{2}, G_{6},  G^{n-1}\Big)+
n(n-1)Str\Big(\zeta_{2},  G^2_{4},  G^{n-2}\Big)
\eeqa
and there are no anomalies with respect to the standard gauge transformations.

{\it In summary, we have the expressions (\ref{Y2n+8}) for primary form $\Upsilon_{2n+8}$,
the expression (\ref{secondary3}) for the secondary form $\rho_{2n+6}$ and
(\ref{amomaly3}) for the anomalies.}

\section{\it Conclusion }

In this article we are interested in enumerating and classifying
{\it metric independent, gauge invariant and closed forms} in generalized YM theory.
The forms that we constructed are
defined in various dimensions, are based on non-Abelian tensor gauge fields and
are polynomial on the corresponding fields strength tensors - curvature forms.
All these forms $\Phi_{2n+4} $, $\Xi_{2n+6}$ and $\Upsilon_{2n+8}$ are analogous to the
Pontryagin-Chern-Simons densities $\CP_{2n}$  in YM gauge theory (\ref{abeliananom}).
They are closed forms, but not globally exact.

The secondary characteristic classes $\psi_{2n+3}$,  $\phi_{2n+5}$
and $\rho_{2n+7}$  have been expressed in integral form
(\ref{integral}),(\ref{secondaryform2})  and (\ref{seccondary3}) in analogy with the
Chern-Simons form (\ref{celebratedanomaly}).
The secondary forms are not unique, because they
can be modified by the addition of the differential
of a one-step-lower-order forms. By adding the properly chosen exact forms
(\ref{ready}), (\ref{secondary2}) and (\ref{secondary3})
respectively to the secondary forms
$\psi_{2n+3}$,  $\phi_{2n+5}$ and
$\rho_{2n+7}$, we are led to much more simple expressions.
The gauge variation of the secondary forms can also be found: (\ref{amomaly0}), (\ref{amomaly2})
and (\ref{amomaly3})
yielding the potential anomalies in gauge field theory. The above general considerations
should be supplemented  by an explicit calculation of loop diagrams involving chiral fermions.
The argument in favor of the existence of
these potential anomalies is based on the fact that they fulfill Wess-Zumino
consistency conditions \cite{Wess:1971yu,Zumino:1983ew,Stora:1983ct,Faddeev:1984jp,
Faddeev:1985iz,AlvarezGaume:1985ex}.
The integrals of these forms over the corresponding space-time coordinates
provide us with new topological Lagrangians \cite{Antoniadis:2013jja} and with a generalization of the
Chern-Simons quantum field theory
\cite{ Witten:1988hf,Schwarz:1978cm,Schwarz:1978cn,Schonfeld:1980kb,Deser:1982vy,
Deser:1981wh,Witten:1992fb, Beasley:2005vf, Witten:1989ip,
Axelrod:1989xt, Dijkgraaf:1989pz,Birmingham:1991ty,Blau:1989bq,Blau:1989dh,Thompson:1992hv}.

At the same time, these densities constructed on a high-dimensional manifold
have their own  value.  Their integrals
represent  global geometric invariants suggesting the existence of  new
topological   characterization   of the  manifolds
\cite{Zumino:1983ew,Stora:1983ct,LBL-16443,AlvarezGaume:1983cs,AlvarezGaume:1985yb}.

\section*{\it Acknowledgements}
G.S. would like to thank Ignatios Antoniadis, Ludwig Faddeev and Luis Alvarez-Gaume
for  discussions. This work was supported in part by the
General Secretariat for Research and Technology of Greece and
the European Regional Development Fund under the contract
MIS-448332-ORASY (NSRF 2007-13 ACTION, KRIPIS).

\section{{\it Appendix}}

The gauge transformations of non-Abelian tensor gauge fields
were defined in \cite{Savvidy:2005fi,Savvidy:2005zm,Savvidy:2005ki}:
\beqa\label{gauge trasformation of connections}
\delta A&=&  D \xi  ,  \\
\delta A_3&=& D \zeta_2 + [A_3, \xi]  \nonumber\\
\delta A_5&=& D \zeta_4 + 2[A_3 ,\zeta_2]+  [A_5, \xi] ,\nn\\
\delta A_7&=& D \zeta_6 + 3[A_3 ,\zeta_4]+ 3 [A_5, \zeta_2] + [A_7,\xi] ,\nn\\
\delta A_9&=& D \zeta_8 + 4[A_3 ,\zeta_6]+ 6 [A_5, \zeta_4] +4[A_7, \zeta_2] + [A_9,\xi] ,\nn\\
...........&&..................................., \nn
\eeqa
where $DA_{2n+1}=dA_{2n+1}+\{A,A_{2n+1}\}$ and the corresponding field-strength tensors are
\beqa\label{field-strength tensors}
G  &=& d A  + A^{2} = DA- A^{2} \\
 G_4 &=& d A_3 +\{ A  , A_3 \}= DA_{3},\nn\\
G_6 &=&d A_5 + \{A, A_5\} + \{A_3, A_3\}= DA_{5}+\{A_{3}, A_{3}\},\nn\\
G_8 &=& d A_7 + \{A, A_7\} +3 \{A_3, A_5\}=DA_{7}+3\{A_{3}, A_{5}\},\nn\\
G_{10} &=& dA_9 + \{ A, A_9 \} +4 \{ A_3, A_7\}+ 3 \{A_5, A_5\},\nn\\
...........&&................................... \nn
\eeqa
The general variations of the field-strength tensors are:
\beqa\label{general variation of field strength tensors}
\delta G &=& D (\delta A),\\
\delta G_4 &=& D(\delta A_3) + \{A_3,\delta A \},\nn\\
\delta G_6 &=& D(\delta A_5) + \{A_5,\delta A \}+ 2\{A_3,\delta A_3 \},\nn\\
\delta G_8 &=& D(\delta A_7) + \{A_7,\delta A \}+ 3\{A_5,\delta A_3 \}+ 3\{A_3,\delta A_5 \},\nn\\
\delta G_{10} &=& D(\delta A_9) + \{A_9,\delta A \}+
4\{A_7,\delta A_3 \}+ 6\{A_5,\delta A_5 \}+ 4\{A_3,\delta A_7 \},\nn\\
..........&=&...................................\nn
\eeqa
The gauge transformations
of the field-strength tensors follow from (\ref{general variation of field strength tensors})
and (\ref{gauge trasformation of connections}). They are homogeneous:
\beqa\label{gauge trasnformations of field strength tensors}
\delta G &=& [G,\xi],\\
\delta G_4 &=& [G_4,\xi] +[G,\zeta_2],\nn\\
\delta G_6 &=& [G_6,\xi] + 2[G_4,\zeta_2] +[G,\zeta_4],\nn\\
\delta G_8 &=& [G_8,\xi] + 3[G_6,\zeta_2] +3[G_4,\zeta_4]+[G,\zeta_6],\nn\\
\delta G_{10} &=& [G_{10},\xi] + 4[G_8,\zeta_2] +6[G_6,\zeta_4]+4 [G_4,\zeta_6] +[G,\zeta_8],\nn\\
...........&&................................... \nn
\eeqa
The Bianchi identities  are  given by
\beqa\label{Bianchi Identities}
D G  &=& 0,\\
DG_4 + [A_3, G]&=&0,\nn\\
D G_6 +2[A_3, G_4] + [A_5,G] &=&0,\nn\\
D G_8 +3 [A_3, G_6] + 3 [A_5,G_4]+ [A_7,G] &=&0,\nn\\
D G_{10} +4 [A_3, G_8] + 6 [A_5,G_6]+ 4 [A_7,G_4]+[A_9,G]  &=&0,\nn\\
.........................................&&....., \nn
\eeqa
where $DG_{2n}  = dG_{2n}  + [A , G_{2n} ]$. Generalizing  Zumino's construction \cite{Zumino:1983ew},
we introduce a one-parameter family of potentials and field-strengths as :
\beqa\label{Zumino pontentials - connections}
A_{t}= t A,~~~A_{3t}= t A_3,~~~A_{5t}= t A_5,~~~A_{7t}= t A_7,~~~A_{9t}= t A_9,\nn\\
G_{t}= t G +(t^2-t)A^{2},\nn\\
G_{4t}= t G_4 +(t^2-t)\{A,A_{3}\},\nn\\
G_{6t}= t G_6 +(t^2-t)(\{A,A_{5}\}+ \{A_3,A_{3}\}) ,\nn\\
G_{8t}= t G_8 +(t^2-t)(\{A,A_{7}\}+ 3\{A_3,A_{5}\}) ,\nn\\
G_{10t}= t G_{10} +(t^2-t)(\{A,A_{9}\}+ 4\{A_3,A_{7}\}+ 3\{A_5,A_{5}\} ) ,\\
............................................. \nn
\eeqa
The Bianchi identities hold for the deformed fields as well.
\beqa\label{derivatives of zumino tensors}
{\partial G_{t}\over \partial t}&=&dA+2tA^{2}=D_{t}A,\\
{\partial G_{4t}\over \partial t}&=&dA_{3}+2t\{A, A_{3}\}=D_{t}A_{3}+t\{A, A_{3}\},\nn\\
{\partial G_{6t}\over \partial t}&=&dA_{5}+2t\Big(\{A, A_{5}\}+\{A_{3}, A_{3}\}\Big)=D_{t}A_{5}+t\{A, A_{5}\}+2t\{A_{3}, A_{3}\},\nn\\
{\partial G_{8t}\over \partial t}&=&dA_{7}+2t\Big(\{A, A_{7}\}+3\{A_{3}, A_{5}\}\Big)= D_{t}A_{7}+t\{A, A_{7}\}+6t\{A_{3}, A_{5}\},\nn
\eeqa
where $D_t A_{2n+1}= d A_{2n+1}+\{A_t , A_{2n+1}\}$.
Because we  used the properties  (B.10) of the symmetrized traces which are
defined in  \cite{LBL-16443} we shall present them in the form convenient for our
purposes:
\beqa\label{B10}
\sum^{n}_{i=1}(-1)^{(d_1+...+d_{i-1})d_{\theta}} Str(\Lambda_1,...,
[\Theta, \Lambda_i],...\Lambda_n)=0,~~~~~~~~~\nn~~~~~~~~~~~~~~~~~~~~~~~(B.10)
\eeqa
where $d_{i}$ is the rank of the form $\Lambda_i$ and $\Theta$ is an even
form. But if both $\Theta$ and $\Lambda_i$ are odd forms, then  the commutator should
be replaced by the anticommutator. For the exterior derivative we use \cite{LBL-16443}:
\beqa\label{B13}
d Str(\Lambda_1,... , \Lambda_i ,...\Lambda_n) =\sum^{n}_{i=1}(-1)^{d_1+...+d_{i-1} }
Str(\Lambda_1,...,  D \Lambda_i ,...\Lambda_n).\nn~~~~~~~~~~~~~~~~~(B.13)
\eeqa

\vfill

\begin{thebibliography}{99}

\bibitem{Adler:1969gk}
  S.~L.~Adler,
\emph{Axial vector vertex in spinor electrodynamics,}
 Phys.\ Rev.\  {\bf 177} (1969) 2426.  

\bibitem{Bell:1969ts}
  J.~S.~Bell and R.~Jackiw,
\emph{A PCAC puzzle: $\pi_0$ to $ \gamma \gamma$ in the sigma model,}
  Nuovo Cim.\  A {\bf 60} (1969) 47.

\bibitem{Bardeen:1969md}
  W.~A.~Bardeen,
\emph{Anomalous Ward identities in spinor field theories,}
  Phys.\ Rev.\  {\bf 184} (1969) 1848.  

\bibitem{Wess:1971yu}
  J.~Wess and B.~Zumino,
\emph{Consequences of anomalous Ward identities,}
Phys.\ Lett.\ B {\bf 37} (1971) 95.  


\bibitem{Frampton:1983nr}
  P.~H.~Frampton and T.~W.~Kephart,
\emph{The Analysis Of Anomalies In Higher Space-time Dimensions,}
  Phys.\ Rev.\ D {\bf 28} (1983) 1010.  

\bibitem{Zumino:1983ew}
  B.~Zumino,
\emph{Chiral Anomalies And Differential Geometry}, Lectures Given At Les Houches,
  August 1983, LBL-16747, UCB-PTH-83/16




\bibitem{Stora:1983ct}
  R.~Stora,
\emph{Algebraic Structure And Topological Origin Of Anomalies,}
LAPP-TH-94, Nov 1983. 20pp.
Seminar given at Cargese Summer Inst.: Progress in Gauge Field Theory, Cargese, France, Sep 1-15, 1983.
Published in Cargese Summer Inst.1983:0543

\bibitem{Faddeev:1984jp}
  L.~D.~Faddeev,
\emph{Operator Anomaly For The Gauss Law,}
  Phys.\ Lett.\  B {\bf 145} (1984) 81.

\bibitem{Faddeev:1985iz}
  L.~D.~Faddeev and S.~L.~Shatashvili,
\emph{Algebraic and Hamiltonian Methods in the Theory of Nonabelian Anomalies,}
  Theor.\ Math.\ Phys.\  {\bf 60} (1985) 770   [Teor.\ Mat.\ Fiz.\  {\bf 60} (1984) 206].  




\bibitem{LBL-16443}
  B.~Zumino, Y.~-S.~Wu and A.~Zee,
\emph{Chiral Anomalies, Higher Dimensions, and Differential Geometry,}
  Nucl.\ Phys.\ B\ {\bf 239}, 477  (1984).  

\bibitem{Manes:1985df}
  J.~Manes, R.~Stora and B.~Zumino,
\emph{Algebraic Study Of Chiral Anomalies,}
  Commun.\ Math.\ Phys.\  {\bf 102} (1985) 157.



\bibitem{Treiman:1986ep}
  S.~B.~Treiman, E.~Witten, R.~Jackiw and B.~Zumino,
\emph{Current Algebra And Anomalies,}
{\it  Singapore, Singapore: World Scientific ( 1985) 537p}

\bibitem{Faddeev:1987hg}
  L.~D.~Faddeev,
\emph{Hamiltonian approach to the theory of anomalies,}
{  in *Schladming 1987, Proceedings, Recent Developments In Mathematical Physics* 137-159. }

\bibitem{Faddeev:1986pc}
  L.~D.~Faddeev and S.~L.~Shatashvili,
\emph{Realization of the Schwinger Term in the Gauss Law and the Possibility
of Correct Quantization of a Theory with Anomalies,}
  Phys.\ Lett.\ B {\bf 167} (1986) 225.  


\bibitem{DKFaddeev}
  D.~K.~Faddeev,
\emph{The operation $\delta$ is the  coboundary operator in homological algebra.}
Dokl.\ Akad.\ Nauk\ SSSR, {\bf 8} (1947) 361.

\bibitem{AlvarezGaume:1983cs}
  L.~Alvarez-Gaume and P.~H.~Ginsparg,
\emph{The Topological Meaning of Nonabelian Anomalies,}
  Nucl.\ Phys.\ B {\bf 243} (1984) 449.  


\bibitem{AlvarezGaume:1985yb}
  L.~Alvarez-Gaume and P.~H.~Ginsparg,
\emph{Geometry Anomalies,}
  Nucl.\ Phys.\ B {\bf 262} (1985) 439.  


\bibitem{AlvarezGaume:1985ex}
  L.~Alvarez-Gaume,
\emph{An Introduction To Anomalies,} HUTP-85/A092.  
Lectures given at the International School of Mathematical Physics on
{\it Fundamental Problems of Gauge Field Theory},  Erice, Italy,1-14 July 1985.





\bibitem{arXiv:1001.2808}
  G.~Savvidy,
\emph{Topological mass generation in four-dimensional gauge theory,}
Phys.\ Lett.\ B\ {\bf 694}, 65  (2010)  [arXiv:1001.2808 [hep-th]].

\bibitem{Antoniadis:2012ep}
  I.~Antoniadis and G.~Savvidy,
\emph{New gauge anomalies and topological invariants in various dimensions,}
  Eur.\ Phys.\ J.\ C {\bf 72} (2012) 2140  [arXiv:1205.0027 [hep-th]].  


\bibitem{Antoniadis:2013jja}
  I.~Antoniadis and G.~Savvidy,
\emph{Extension of Chern-Simons forms and new gauge anomalies,}
 Int.\ J.\ Mod.\ Phys.\ A {\bf 29} (2014) 401  [arXiv:1304.4398 [hep-th]].



\bibitem{Savvidy:2005fi}
G.~Savvidy,
\emph{Non-Abelian tensor gauge fields: Generalization of Yang-Mills theory,}
Phys.\ Lett.\ B {\bf 625} (2005) 341
[arXiv:hep-th/0509049]


\bibitem{Savvidy:2005zm}
  G.~Savvidy,
 \emph{Non-abelian tensor gauge fields. I,}
  Int.\ J.\ Mod.\ Phys.\ A {\bf 21} (2006) 4931.

\bibitem{Savvidy:2005ki}
  G.~Savvidy,
  \emph{Non-abelian tensor gauge fields. II,}
  Int.\ J.\ Mod.\ Phys.\ A {\bf 21} (2006) 4959.



\bibitem{Witten:1988hf}
  E.~Witten,
 \emph{Quantum Field Theory and the Jones Polynomial,}
 Commun.\ Math.\ Phys.\  {\bf 121} (1989) 351.


\bibitem{Schwarz:1978cm}
  A.~S.~Schwarz,
 \emph{On Quantum Fluctuations of Instantons,}  Lett.\ Math.\ Phys.\  {\bf 2} (1978) 201.  

\bibitem{Schwarz:1978cn}
  A.~S.~Schwarz,
 \emph{The Partition Function of Degenerate Quadratic Functional and Ray-Singer Invariants,} Lett.\ Math.\ Phys.\  {\bf 2} (1978) 247.  

\bibitem{Schonfeld:1980kb}
  J.~F.~Schonfeld,
 \emph{A Mass Term For Three-Dimensional Gauge Fields,}
  Nucl.\ Phys.\  B {\bf 185} (1981) 157.


\bibitem{Deser:1982vy}
  S.~Deser, R.~Jackiw and S.~Templeton,
\emph{Three-Dimensional Massive Gauge Theories,}
  Phys.\ Rev.\ Lett.\  {\bf 48} (1982) 975.

\bibitem{Deser:1981wh}
  S.~Deser, R.~Jackiw and S.~Templeton,
\emph{Topologically massive gauge theories,}
  Annals Phys.\  {\bf 140} (1982) 372




\bibitem{Witten:1992fb}
  E.~Witten,
 \emph{Chern-Simons gauge theory as a string theory,}  Prog.\ Math.\  {\bf 133} (1995) 637  [hep-th/9207094].  

\bibitem{Beasley:2005vf}
  C.~Beasley and E.~Witten,
 \emph{Non-Abelian localization for Chern-Simons theory,}  J.\ Diff.\ Geom.\  {\bf 70} (2005) 183  [hep-th/0503126].  

\bibitem{Witten:1989ip}
  E.~Witten,
 \emph{Quantization Of Chern-simons Gauge Theory With Complex Gauge Group,}  Commun.\ Math.\ Phys.\  {\bf 137} (1991) 29.  


\bibitem{Axelrod:1989xt}
  S.~Axelrod, S.~Della Pietra and E.~Witten,
 \emph{Geometric Quantization Of Chern-simons Gauge Theory,}  J.\ Diff.\ Geom.\  {\bf 33} (1991) 787.  


\bibitem{Dijkgraaf:1989pz}
  R.~Dijkgraaf and E.~Witten,
 \emph{Topological Gauge Theories and Group Cohomology,}  Commun.\ Math.\ Phys.\  {\bf 129} (1990) 393.  

\bibitem{Birmingham:1991ty}
  D.~Birmingham, M.~Blau, M.~Rakowski and G.~Thompson,
 \emph{Topological field theory,}  Phys.\ Rept.\  {\bf 209} (1991) 129.  

\bibitem{Blau:1989bq}
  M.~Blau and G.~Thompson,
 \emph{Topological Gauge Theories of Antisymmetric Tensor Fields,}  Annals Phys.\  {\bf 205} (1991) 130.  

\bibitem{Blau:1989dh}
  M.~Blau and G.~Thompson,
 \emph{A New Class Of Topological Field Theories And The Ray-singer Torsion,}  Phys.\ Lett.\ B {\bf 228} (1989) 64.  


\bibitem{Thompson:1992hv}
  G.~Thompson,
 \emph{1992 Trieste lectures on topological gauge theory and Yang-Mills theory,}  In *Trieste 1992, Proceedings, High energy physics and cosmology* 1-75 and Trieste Int. Cent. Theor. Phys. - IC-93-112 (93/05,rec.Jul.) 75 p. (310578) (see Conference Index)  [hep-th/9305120].  











\end{thebibliography}
\end{document}